# PT-Invariance and Indefinite Metric

Scott Chapman
*Chapman University, One University Drive, Orange, CA 92866*
(Dated: October 16, 2009)

**Abstract**

A new proof is given for why the non-Hermitian, PT-Invariant cubic oscillator with imaginary coupling has real eigenvalues. The proof consists of two steps. In the first step, it is shown that for many PT-Invariant Hamiltonians, one can define corresponding Hermitian Hamiltonians that have the same eigenvalues when quantized using indefinite metric. The second step is to show that the indefinite-metric eigenvalues of the Hermitian cubic oscillator are real since the norms of its eigenstates do not vanish. The correspondence between PT-Invariant Hamiltonians and indefinite-metric Hermitian Hamiltonians is further discussed as a way to determine vacuum stability for certain supersymmetric gauge theories.

PACS numbers: 12.10.-g, 12.60.Jv

Several years ago Bessis conjectured that all of the eigenvalues of the cubic oscillator with purely imaginary coupling constant were real [1]. This was surprising, since the Hamiltonian in this theory is not Hermitian, but just invariant under parity (P) and time reversal (T). Later Bender and Boettcher showed numerically that an entire class of nonHermitian, PT-invariant Hamiltonians has only real, positive eigenvalues [2]. Since then, PT-invariant theories have been extensively studied [3-10], and analytic proofs of Bessis' conjecture in various limits have been provided [4,6].

The purpose of this paper is to provide a new proof of Bessis' conjecture that illuminates a connection between PT-invariance and indefinite metric. The first step is to show that for a class of PT-invariant Hamiltonians, one can define corresponding Hermitian Hamiltonians that have the same eigenvalues when quantized with indefinite metric. The second step is to show that the specific Hermitian Hamiltonian corresponding to the PT-Invariant cubic oscillator has real eigenvalues when quantized using indefinite metric. A key element of this second step is to show that no eigenstate of the Hermitian Hamiltonian has vanishing norm. This condition is similar to a condition presented in [6] in the context of a more general, complex metric space.

Consider the following class of PT-invariant Hamiltonians:

$$\overline{H}_k = \omega\left(p^2 + x^2 - g(ix)^k\right), \qquad (1)$$

where $k$ is a positive integer, $g$ is a real constant, and $x$ and $p$ are Hermitian operators that satisfy $[p, x] = -i$. For every PT-invariant Hamiltonian of the form $\overline{H}_k$, one can define a corresponding Hermitian Hamiltonian that has an overall "wrong sign":

$$H_k = -\omega\left(p^2 + x^2 + gx^k\right). \qquad (2)$$

It is well known that if a Hamiltonian is quantized using indefinite metric, the Hermitian operators $x$ and $p$ have imaginary eigenvalues [11]. In this case, it can be helpful to replace the standard operators with anti-Hermitian operators that have real eigenvalues [12]:

$$\tilde{x} \equiv -ix \qquad \tilde{p} \equiv ip. \qquad (3)$$

Re-expressing the Hermitian Hamiltonian $H_k$ in terms of these ("indefinite-metric") operators,

$$H_k = \omega\left(\tilde{p}^2 + \tilde{x}^2 - g(i\tilde{x})^k\right), \qquad (4)$$

the Hermitian Hamiltonian $H_k$ takes exactly the same functional form as the PT-Invariant Hamiltonian $\overline{H}_k$ (by construction). Since the position and momentum operators in both $H_k$ and $\overline{H}_k$ have real eigenvalues and the same commutation relations $[p, x] = [\tilde{p}, \tilde{x}] = -i$, it follows that both Hamiltonians have the same energy eigenvalues. This will be shown below in a different way by using the holomorphic representation.

To quantize $\overline{H}_k$ in a positive-definite metric space, one may define ladder operators

$$a \equiv \frac{1}{\sqrt{2}}(x + ip) \qquad \text{and} \qquad a^\dagger \equiv \frac{1}{\sqrt{2}}(x - ip) \qquad (5)$$

that have the usual commutation relations

$$[a, a^\dagger] = 1. \qquad (6)$$

To build a complete set of quantum states with positive-definite metric, one may start by defining $|0\rangle$ through

$$a|0\rangle = 0 \qquad \langle 0\|0\rangle = 1. \qquad (7)$$

Additional states can then be obtained by applying any number of creation operators to $|0\rangle$:

$$|n\rangle \equiv \frac{1}{n!}(a^\dagger)^n |0\rangle. \tag{8}$$

Since this is a complete basis, any eigenstate of the Hamiltonian can be expanded in this basis. For example, let the expansion of the eigenstate $|\phi_m\rangle$ be given by

$$|\phi_m\rangle = \sum_n c_{mn} |n\rangle, \tag{9}$$

where $c_{mn}$ are complex constants.

A similar quantization process can be carried out for the Hermitian Hamiltonian $\overline{H}_k$, except using an indefinite-metric space of states. Namely, one can define

$$\tilde{a} \equiv \frac{1}{\sqrt{2}}(\tilde{x} + i\tilde{p}) \quad \text{and} \quad (-\tilde{a}^\dagger) \equiv \frac{1}{\sqrt{2}}(\tilde{x} - i\tilde{p}). \tag{10}$$

The minus sign in front of $\tilde{a}^\dagger$ reflects the fact that $\tilde{x}$ and $\tilde{p}$ are anti-Hermitian. Using these operators, the commutation relation analogous to (6) is

$$[\tilde{a}, (-\tilde{a}^\dagger)] = 1. \tag{11}$$

One can similarly define $\tilde{a}|\tilde{0}\rangle = 0$ and a complete basis of states by

$$|\tilde{n}\rangle \equiv \frac{1}{n!}(-\tilde{a}^\dagger)^n |\tilde{0}\rangle. \tag{12}$$

This basis has indefinite metric since states with odd $n$ have negative norms, as can be seen from the commutation relation (11).

From (1), (4), (5), and (10), $H_k$ and $\overline{H}_k$ have exactly the same functional form, except that $\tilde{a}$ replaces $a$ and $(-\tilde{a}^\dagger)$ replaces $a^\dagger$. In the context of these replacements, the operator commutation relations (6) and (11) are the same, and the bases of states (8) and (12) are the same. Therefore it follows that if one defines

$$|\tilde{\phi}_m\rangle \equiv \sum_n c_{mn} |\tilde{n}\rangle, \tag{13}$$

then

$$\overline{H}_k |\phi_m\rangle = E_m |\phi_m\rangle \quad \text{if and only if} \quad H_k |\tilde{\phi}_m\rangle = E_m |\tilde{\phi}_m\rangle \tag{14}$$

In other words, $\bar{H}_k$ must have the same eigenvalues as $H_k$ quantized with indefinite metric.

The next question is whether the Hermitian Hamiltonian $H_k$ still has real eigenvalues when it is quantized with indefinite metric. For a Hermitian Hamiltonian, one finds:

$$\langle \tilde{\phi}_m | H_k | \tilde{\phi}_m \rangle = E_m \langle \tilde{\phi}_m \| \tilde{\phi}_m \rangle = E^*_m \langle \tilde{\phi}_m \| \tilde{\phi}_m \rangle, \tag{15}$$

where the first equality is from $H_k$ acting to the right and the second is from $H_k$ acting to the left. Therefore, if $\langle \tilde{\phi}_m \| \tilde{\phi}_m \rangle \neq 0$, then the eigenvalue $E_m$ is real. Since indefinite metric spaces do have states with zero norm, one cannot automatically conclude that Hermitian Hamiltonians have real eigenvalues when quantized with indefinite metric. More particularly, if any eigenstate of the Hamiltonian has zero norm, then the corresponding eigenvalue is not necessarily real. On the other hand, if one can prove that all of the eigenstates of $H_k$ have non-zero norms, then it follows that all of its eigenvalues are real.

Gell-Mann and Low developed a method for determining eigenstates of an interacting Hamiltonian $H = H_0 + gH_I$ in terms of an exactly solvable free Hamiltonian $H_0$ [13]. Namely, for every state $|\tilde{n}\rangle$ that is an eigenstate of $H_0$, one may define a state $|\tilde{\phi}_n\rangle$ that is an eigenstate of $H = H_0 + gH_I$ as follows:

$$|\tilde{\phi}_n\rangle \equiv \lim_{\varepsilon \to 0} \frac{U_\varepsilon |\tilde{n}\rangle}{\langle \tilde{n} | U_\varepsilon | \tilde{n} \rangle} \tag{16}$$

$$U_\varepsilon = \sum_{n=0}^{\infty} (-ig)^n \int_{-\infty}^{0} dt_1 \int_{-\infty}^{t_1} dt_2 \cdots \int_{-\infty}^{t_{n-1}} dt_n\, e^{\varepsilon(t_1+\ldots+t_n)} H_I(t_1) \cdots H_I(t_n) \tag{17}$$

$$H_I(t) \equiv \exp(iH_0 t) H_I \exp(-iH_0 t)$$
$$= -\omega\left[ \tilde{a}^{\dagger 3} e^{3i\omega t} + \tilde{a}^3 e^{-3i\omega t} + 3(\tilde{a}^{\dagger 2}\tilde{a} - \tilde{a}^\dagger)e^{i\omega t} + 3(\tilde{a}^\dagger \tilde{a}^2 - \tilde{a})e^{-i\omega t} \right], \tag{18}$$

where the last expression in (18) is specific to the cubic oscillator $H_3$.

After integrating over $t_n$, a term in $H_I(t_n)$ with time dependence $\exp(im\omega t_n)$ will combine with a term in $H_I(t_{n-1})$ with time dependence $\exp(im'\omega t_{n-1})$ to give an overall oscillatory time dependence $\exp(i(m+m')\omega t_{n-1})$ in the $t_{n-1}$ integral. If $m = -m'$, then the $t_{n-1}$ integral for that term will be proportional to $\varepsilon^{-1}$. It is straightforward to see that the only way to get a factor of $\varepsilon^{-1}$ in a $dt_k$ integral in (16) is if the combined oscillatory factors from earlier times cancel. This can only happen for contributions from $H_I(t_k) \cdots H_I(t_n)$ that have an equal number of creation and destruction operators. More generally, for each term with $\varepsilon^{-n}$ dependence, there will be some function of the number operator

$$\tilde{N} = -\tilde{a}^\dagger \tilde{a}, \tag{19}$$

all the way to the right of the term.

Keeping track of factors of $i$, and considering that $U_\varepsilon$ is unitary in the $\varepsilon \to 0$ limit, one finds that to any order of perturbation theory

$$U_\varepsilon \to \exp(if_1(\tilde{a}, \tilde{a}^\dagger)) \exp\left(\frac{i\omega}{\varepsilon} f_2(\tilde{N})\right) \quad \text{as} \quad \varepsilon \to 0, \tag{20}$$

where $f_1$ and $f_2$ are Hermitian functions involving $g$ but independent of $\varepsilon$. The existence of the infinite phase in $U_\varepsilon$ is why Gell-Mann and Low included the denominator in the definition of $|\tilde{\phi}_n\rangle$ in (16) – so that the infinite phase in the numerator is cancelled by the denominator. The norm of each eigenstate is therefore given by:

$$\langle \tilde{\phi}_n \| \tilde{\phi}_n \rangle = \frac{\langle \tilde{n} \| \tilde{n} \rangle}{|\langle \tilde{n} | \exp(if_1) | \tilde{n} \rangle|^2} \neq 0 \tag{21}$$

Since none of the norms of eigenstates vanish, it follows from (14) that all of the eigenvalues of $H_3$ are real, and therefore that all of the eigenvalues of the PT-Invariant cubic oscillator $\overline{H}_3$ are real.

For the purpose of proving that PT-invariant eigenvalues are real, it does not matter that half of the norms in (21) are negative – it only matters that they are not zero. However, if one was interested in exploring the physics of the Hamiltonian $H_k$, it *would* matter. To have a sensible theory, one would have to remove all of the negative norm states. Fortunately, the Gell-Mann and Low formalism provides a consistent way to do this. One could restrict the "physical" basis of states to $|\tilde{\phi}_n\rangle$ (as defined by (16)) with $n$ even. This is a consistent condition, because the Hamiltonian acting on any linear combination of physical states would also be a physical state, since $|\tilde{\phi}_n\rangle$ are eigenstates.

One could call the Hamiltonian $H_k$ a "wrong-sign" Hamiltonian since it has a kinetic energy term $-p^2$ with the wrong sign. In standard positive-definite-metric quantization, a theory like this would not have a stable vacuum, since one could always add more particles to any state to reduce the energy. But using indefinite metric and keeping in mind the correspondence between indefinite-metric quantization and PT-invariant theories, one can see that wrong-sign Hamiltonians like $H_k$ can have stable vacuums. Combining this with the method suggested here for removing negative-norm states, these wrong-sign theories can have consistent physical and probabilistic interpretations.

For example, consider "wrong-sign" $\phi^4$ theory:

$$H = -\int d^3x \left( \tfrac{1}{2} P^2 + \tfrac{1}{2}(\nabla\phi)^2 + \tfrac{1}{2}m^2\phi^2 + \lambda\phi^4 \right). \tag{22}$$

where $P = -\partial_0 \phi$. In standard quantization, this theory has negative-definite energy and no stable vacuum. But using indefinite-metric quantization one has

$$H = \int d^3x \left( \tfrac{1}{2} \widetilde{P}^2 + \tfrac{1}{2}(\nabla\widetilde{\phi})^2 + \tfrac{1}{2}m^2\widetilde{\phi}^2 - \lambda\widetilde{\phi}^4 \right), \tag{23}$$

where $\widetilde{\phi} \equiv -i\phi$ and $\widetilde{P} \equiv iP$. Generalizing the arguments presented earlier to field theory, this Hamiltonian has the same eigenvalues as its corresponding PT-Invariant Hamiltonian – one that has been found to have a stable vacuum for Euclidean dimensions < 3 [14,15]. Even more interestingly, in some cases this theory develops an *imaginary* vacuum expectation value for $\widetilde{\phi}$. In indefinite metric quantization, the fact that this VEV is imaginary rather than real makes intuitive sense since $\widetilde{\phi}$ is anti-Hermitian. Indefinite metric would provide its operator piece with real eigenvalues, but a constant piece would have to be imaginary.

The correspondence between PT-invariance and indefinite metric is also interesting for supersymmetric gauge theories like SU(2/1) [16-19] and SU(2/3) that have "wrong sign" sectors. For example in SU(2/3), the bosonic gauge field subgroup is U(1) x SU(2) x SU(3), but the SU(3) gauge field has the "wrong sign" [20]:

$$\mathcal{L} = \tfrac{1}{4} F^a_{\mu\nu} F^{a\mu\nu}. \tag{24}$$

When quantized using indefinite metric, this "wrong-sign QCD" takes the form of a PT-Invariant theory with imaginary cubic coupling and a negative quartic term – like the one in (23) that leads to an imaginary VEV. Making a connection with the corresponding PT-invariant theory introduces the possibility that (24) could have a stable vacuum. Perhaps it could also generate some vacuum expectation value for a gluon condensate $F^a_{\mu\nu}F^{\mu\nu a}$. If so, this could lead to a confining vacuum energy density. The restriction to only positive norm states in wrong-sign QCD would also limit the types of gluon states that could be observed experimentally.

A new proof has been given that certain types of non-Hermitian, PT-Invariant Hamiltonians have real eigenvalues. The proof consists of mapping these Hamiltonians to Hermitian counterparts with indefinite metric, then showing that the Hermitian counterparts have real eigenvalues. It is hoped that the techniques used may be helpful both in further studies of PT-Invariant Hamiltonians and also in studies of indefinite-metric, Hermitian Hamiltonians like those that arise in certain supersymmetric gauge theories.